\documentclass[useAMS,usenatbib,usegraphicx,epsfig]{aa}
\usepackage{txfonts,graphicx,epsfig}
\def\eps@scaling{.95}
\def\epsscale#1{\gdef\eps@scaling{#1}}
\def\plotone#1{\centering \leavevmodes
\epsfxsize=\eps@scaling\columnwidth \epsfbox{#1}}

\def\micron{\mu {\rm m}}
\begin{document}

\title{Monte Carlo radiative transfer in SPH density fields}
\titlerunning{Radiative transfer in SPH density fields}

\author{Dimitris Stamatellos \& Anthony~P.~Whitworth}

\offprints{D.~Stamatellos\\ \email{D.Stamatellos@astro.cf.ac.uk}}

\institute{School of Physics \& Astronomy, Cardiff University, 
        5 The Parade, Cardiff CF24 3YB, Wales, UK}

  \date{Received ..., 2005; accepted ... , 2005}

\abstract{
We combine a Monte Carlo radiative transfer code with an SPH 
code, so that -- assuming thermal equilibrium -- we can calculate 
dust-temperature fields, spectral energy distributions, and isophotal maps,  
for the individual time-frames generated by an SPH simulation. On large 
scales, the radiative transfer cells (RT cells) are borrowed from the tree 
structure built by the SPH code, and are chosen so that their size -- and 
hence the resolution of the calculated temperature field -- is comparable 
with the resolution of the density field. We refer collectively to these 
cubic RT cells as the {\it global grid}. The code is tested and found 
to treat externally illuminated dust configurations very well. However, when 
there are embedded discrete sources, i.e. stars, these produce very steep 
local temperature gradients which can only be modelled properly if -- 
in the immediate vicinity of, and centred on, each embedded star -- we 
supplement the global grid with a {\it star grid} of closely spaced 
concentric RT cells.

}
 
\maketitle

\section{Introduction}

Smoothed Particle Hydrodynamics (SPH) (Lucy 1977; Gingold \& Monaghan 
1977) is a Lagrangian computational method which invokes a large array 
of particles to describe a fluid, assigns properties such as mass, 
position and velocity to each particle, and estimates intensive 
thermodynamic variables like density and pressure (and their 
derivatives) using local averages (for reviews see Benz 1990, 1991; 
Monaghan 1992). The main advantage of this method is that, in 
principle, there are no limitations on the geometry of the system or 
how far it can evolve from the initial conditions.

Because of the large computational cost incurred when radiative 
transfer is modelled fully in three dimensions, many previous 
treatments have been limited to one or two dimensions (e.g. 
Efstathiou \& Rowan-Robinson~1990, 1991; Men'shchikov \& 
Henning~1997, Zucconi et al. 2001), and radiative transfer has 
only recently been included in SPH simulations, where a fully 
three-dimensional treatment is essential (Kessel-Deynet \& Burkert 
2000, Oxley \& Woolfson 2003, Whitehouse \& Bate 2004, Viau et 
al. 2005). The Monte Carlo approach for equilibrium radiative 
transfer (Lefevre et al.~1982, 1983; Wolf et al. 1999) is well 
suited to systems with arbitrary geometries, but, as with any 
Monte Carlo method, it is computationally expensive, especially 
when iteration is required. Recently, Bjorkman \& Wood (2001), 
extended an idea by Lucy (1999) and proposed a method to avoid
iteration, by remitting photons as soon as they are absorbed, 
with a frequency distribution adjustment technique. This method 
avoids iteration, and it is fast when compared with the traditional 
Monte Carlo radiative transfer methods. It has been applied 
successfully to a variety of problems, such as protoplanetary discs 
(e.g. Wood et al. 2002), prestellar cores (e.g. Stamatellos et al. 
2004) and Class I objects (e.g. Whitney et al. 2003).

In this paper, we develop and test a method which uses Monte Carlo radiative 
transfer with frequency distribution adjustment, to calculate the 
dust-temperature field, the spectral energy distribution (SED), and 
isophotal maps, for individual time-frames from SPH simulations. We 
use the SPH tree to construct a grid of cubic radiative transfer 
cells (hereafter RT cells) with which the photons interact; we refer 
to this grid as {\it the global grid}. The global grid obtained 
in this way is similar to the tree-structured adaptive grid invoked 
by Kurosawa \& Hillier (2001), but because it borrows the tree 
structure already derived as part of the SPH code, it 
minimizes the computational burden that will be added to the 
hydrodynamic simulations when the two methods are combined. A 
similar method has been used by Oxley \& Woolfson (2003), but they 
adopt an average, wavelength-independent opacity. To account for 
the large temperature gradients that are expected very close to 
stars, we supplement the global grid described above with a grid 
of concentric spherical radiative transfer cells around each star; 
we refer to this supplementary grid as {\it the star grid}.

\section
{Monte Carlo radiative transfer with frequency distribution adjustment}

The  Monte Carlo radiative transfer method (e.g. Stamatellos \& 
Whitworth 2003)  uses a large number of monochromatic luminosity 
packets ($L$-packets) to represent the radiation injected into 
the medium (e.g. stellar radiation, cooling radiation, background 
radiation). Each $L$-packet is given a random frequency, $\nu$, and a 
random optical depth, $\tau$, 
which determines its free path, i.e. how far it travels before 
interacting with the medium. The medium itself is 
divided into a number of RT cells, each with uniform density and 
temperature. Each time an $L$-packet  is scattered, it is given a 
new random direction, using the Henyey \& Greenstein (1941) 
scattering phase function, and a new random optical depth, $\tau$. 
Each time an  $L$-packet is absorbed, it raises the temperature of 
the RT cell in which it is absorbed from $T$ to $T+\Delta T$, and the 
packet is immediately reemitted isotropically, so that the dust is 
in radiative equilibrium. The temperature of the RT cell is computed 
by  equating the total absorbed luminosity to the total emitted 
luminosity. The frequency of the reemitted $L$-packet  is chosen 
using a probability distribution function (PDF) constructed from 
the difference between the emissivity of the cell before and after 
absorption of the  $L$-packet (Bjorkman \& Wood 2001; Baes et al. 
2005),
\begin{eqnarray} \nonumber
p(\nu)\,{\rm d}\nu & = & \frac
{\kappa_\nu\,[B_\nu(T+\Delta T)-B_\nu(T)]\,{\rm d} \nu}
{\int_0^\infty \kappa_\nu\,[B_\nu(T+\Delta T)-B_\nu(T)]\,{\rm d}\nu} \\
 & \simeq & \frac
{\kappa_\nu\,B'(T)\,{\rm d}\nu}
{\int_0^\infty \kappa_\nu\,B'(T)\,{\rm d}\nu} \,.
\label{qBW}
\end{eqnarray}
where $B'(T) \equiv {\rm d}B_\nu / {\rm d}T\,$. At the same time the 
$L$-packet is given a new random optical depth, $\tau$. Using the PDF 
of Eqn. (\ref{qBW}), the correct temperature distribution and spectrum 
of the system are obtained without iteration, once all the $L$-packets have 
been propagated through the medium and escaped.  The method accounts 
for both absorption and scattering, it is robust, it is very accurate 
(see tests by Bjorkman \& Wood 2001, Stamatellos \& Whitworth 2003), 
it can readily be parallized, and it can treat systems with arbitrary 
geometries.

\subsection{Construction of radiative transfer cells}

The radiative transfer cells must be constructed 
so that their linear size is less than, or on the order of, the local 
directional temperature- and density- scale-heights (e.g. $\Delta x 
\la {\rm MIN}\{\,|\,d\ell n[\rho]/dx\,|\,^{-1},\,|\,d\ell n[T]/dx\,|\,^{-1}\}$, 
see Stamatellos \& Whitworth 2003). This ensures that the temperature and 
density do not vary too greatly between adjacent RT cells. The 
construction of the RT cells depends on the specific system under study, 
and, in an hydrodynamic simulation, where the system changes from step 
to step, the RT cells need to be reconstructed at every step. 
Therefore, a robust and efficient algorithm for the construction of 
RT cells is required.

To construct RT cells we take advantage of the fact that our SPH 
code calculates gravitational forces with the aid of a spatial 
tessellation tree (Barnes \& Hut 1986; Hernquist \& Katz 1989). 
The whole computational domain is contained within a cubic cell, 
the root cell, which is then divided into eight smaller cubic 
cells of equal size. If any of these cells contains more than one 
particle, it is subdivided into eight even smaller cells. This 
procedure continues recursively until the smallest cells each contain 
just one particle or no particles at all. The resulting ensemble of 
cells constitutes the SPH tree. In SPH, the tree is 
used so that distant particles do not contribute individually 
to the local gravitational field, but collectively through the cell 
in which they are contained; the tree is also used to find neighbouring 
particles so that smoothed values of thermodynamic variables and their 
derivatives can be evaluated.

The global grid then comprises the smallest set of RT cells 
which (a) spans the entire computational domain (or at least that part 
of it containing matter) and (b) contains fewer 
than $N_{_{\rm MAX}}$ SPH particles. If $N_{_{\rm MAX}}$ is chosen to be 
of the same order as $N_{_{\rm NEIB}}$ (the number of neighbour particles 
used for smoothing purposes), then the RT cells are 
necessarily similar in size to the resolution of the SPH simulation. 
Specifically we choose $N_{_{\rm MAX}}/N_{_{\rm NEIB}} \sim 2$ to 3, so 
that each RT cell has linear size $\sim 2\,h$ (where $h$ is the SPH 
smoothing length) and contains $\sim 0.3\,N_{_{\rm NEIB}}$ particles.

This method of constructing RT cells is  similar to the method of 
Kurosawa \& Hiller (2001) (see also Kurosawa et al. 2004), but it 
is implemented using the Barnes \& Hut (1986) tree structure. Both 
methods produce RT cells with approximately equal 
mass. It is also similar to the method used by Oxley \& Woolfson (2003), 
but they construct RT cells with size $1/3$ to $2/3$ of 
the SPH smoothing length. Our method has 
the advantage that we can readily compute the density in an RT cell 
from the number of particles in the cell and the cell size, whereas 
in the Oxley \& Woolfson method the density must be computed by an 
SPH sum over the neighbouring particles, and this is computationally 
expensive.

The method we use to construct RT cells, 
is implemented relatively easily within SPH and exploits 
information about the tree already calculated for SPH 
purposes. Thus, it does not add much to the code running time. 
It also furnishes us with an efficient procedure for finding 
which RT cell an 
$L$-packet is in. This is critical for the efficiency of the code, 
because as an $L$-packet propagates through the 
computational domain, this search procedure is performed 
many times.

\subsection{$L$-packet propagation and interaction}

The  propagation of $L$-packets through the computational domain 
is performed in small steps, $\delta S$, which gradually decrease 
the optical depth $\tau$ of the $L$-packet until $\tau=0$, 
whereupon the $L$-packet reaches an interaction point. 
This is the most computationally expensive part of the code,  
particularly as there is no analytical expression for the density 
-- and hence the opacity -- along the path of an $L$-packet. Instead, 
the density is determined by identifying the RT cell through which 
the $L$-packet is passing. After each step $\delta S$, we use the 
SPH tree to find which RT cell $i$ the $L$-packet is now in. The 
search starts from the root cell and proceeds to lower level cells, 
until the required RT cell is reached. The search contributes a 
significant fraction of the code running time, and therefore, it 
is essential to choose the step $\delta S$ appropriately. We use
 \begin{equation}
\delta S={\rm MIN}
\left\{\eta_1 \,S_i,\,
\eta_2\, \bar{\ell}_i,
\tau\, \bar{\ell}_i,
\eta_3\, |{\bf r}|
\right\},
\end{equation}
where $\eta_1$, $\eta_2$, $\eta_3$ are constants in the range 
($0.1,1$) and determine the accuracy; $S_i$ is the linear extent 
of RT cell $i$; $\bar{\ell}_i = \kappa_\lambda \rho_i$ is the 
mean-free-path for the $L$-packet ($\kappa_\lambda$ is the 
monchromatic mass-opacity at the wavelength of the $L$-packet, 
and $\rho_i$ is the density in cell $i$); ${\bf r}$ is the 
position of the photon relative to any discrete source (i.e. star).

Kurosawa \& Hiller (2001) adopt a more detailed approach, in which 
they use the path of an $L$-packet to determine where it intercepts 
the boundaries of the RT cells through which it passes. Our method 
is simpler to implement, it propagates the $L$-packets efficiently 
(usually the propagation step is on the order of the local RT cell 
size), and it is accurate (due to the fact that neighbouring RT cells 
have similar densities).

\section{Tests}
 
We consider a spherical cloud with uniform density $n=8.6\times10^4 
{\rm cm}^{-3}$, mass $M=100\,{\rm M}_{\sun}$ and radius 
$R=3.5\times10^4\,{\rm AU}$, and represent it with randomly distributed 
SPH particles. We then test the method of constructing the global 
grid and propagating $L$-packets through the grid in three ways. (i) 
We use the thermodynamic equilibrium test, where the cloud is bathed in 
an isotropic blackbody radiation field with a given temperature 
(Stamatellos \& Whitworth 2003).(ii) We illuminate the cloud with the 
Black (1994) interstellar radiation field (BISRF) and compare our results 
with the results obtained previously using a spherically symmetric grid 
of concentric cells (Stamatellos \& Whitworth 2003). (iii) We embed a 
low-temperature star at the centre of the cloud and again compare the 
results with those obtained previously using a spherically symmetric 
grid of concentric cells (see online appendix). 
We present the results in the following subsections.

\subsection{Cloud bathed in a blackbody radiation field}
 
\begin{figure}[h]
\centerline{
\includegraphics[width=4.4cm]{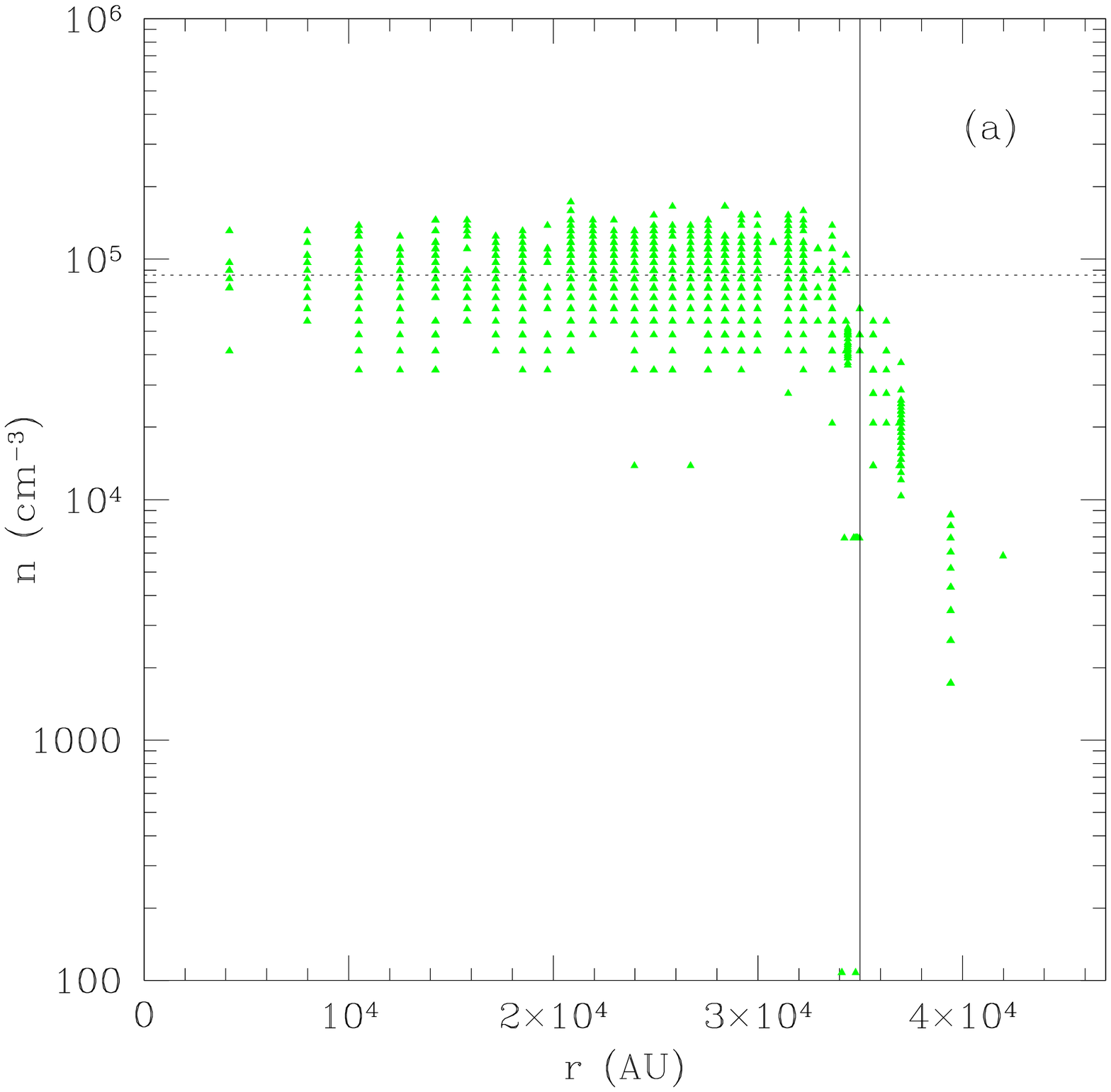}
\includegraphics[width=4.4cm]{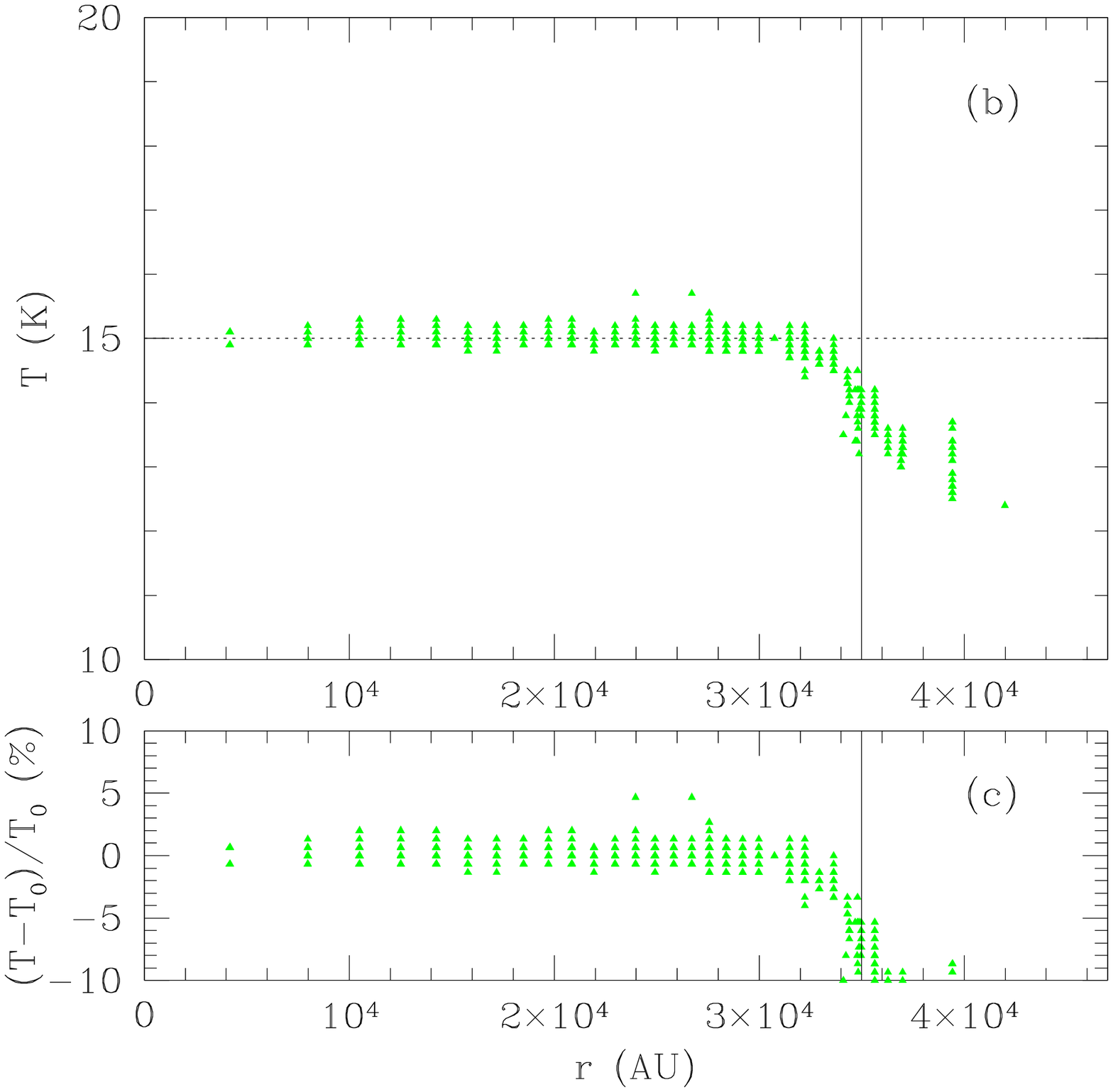}}
\caption{A uniform-density spherical cloud,
represented by 20,000 SPH particles, is illuminated by an isotropic 
blackbody radiation field at $T=15\,{\rm K}$, using $10^8$ $L$-packets. 
RT cells are constructed with $N_{_{\rm MAX}}=60$, giving $N_{_{\rm CELLS}}
=1,484$. {\bf (a)} Triangles give the densities of the individual 
RT cells, the horizontal dotted line marks the mean density, and 
the vertical line marks the boundary of the cloud. {\bf (b)} 
Triangles give the temperatures of the individual RT cells, the 
horizontal dotted line marks $T = 15\,{\rm K}$, and the vertical 
line marks the boundary of the cloud. {\bf (c)} The percentage 
error in the computed temperature.}
\label{fig_sph_thermo9}
\end{figure}

We use $N_{_{\rm TOTAL}} = 20,000$ SPH particles and set the maximum number 
of particles in an RT cell to $N_{_{\rm MAX}} = 60$, giving $N_{_{\rm CELLS}} 
= 1,484$ RT cells, each containing on average ${\bar N} \sim 15$ 
SPH particles. The density in each RT cell is plotted as a function of 
radius in Fig.~\ref{fig_sph_thermo9}a. Because the SPH particles are 
distributed randomly, there are statistical variations in the density 
at the $\pm\,{\bar N}^{-1/2} \simeq \pm\,25\%$ level. There are also 
a number of RT cells that appear to be outside the boundary of the cloud 
and to have very low density. In reality, only a small part of each of 
these RT cells is within the cloud and therefore they contain only a very 
small number of particles. These {\it 
irregular RT cells} are a side effect of representing a spherical cloud 
with a grid of cubic cells. The same problem was also encountered by 
Oxley \& Woolfson (2001).

$10^8$ $L$-packets are then injected from random positions on the 
boundary of the cloud, and with random frequencies and injection 
angles, so as to imitate an isotropic blackbody radiation field at 
$15\,{\rm K}$ (see Stamatellos et al. 2004). The cloud should then 
adopt the temperature of the illuminating radiation field, i.e. 
$15\,{\rm K}$. The computed temperature of each RT cell is plotted 
on Fig.~\ref{fig_sph_thermo9}b. In the body of the cloud the 
temperature is very close to $15\,{\rm K}$, with the error being less 
than $\pm\,3\%$. The errors are higher near and outside the boundary of 
the cloud, because of the presence of the irregular RT cells. These 
errors can be reduced by increasing the number of SPH particles, 
and/or the number of RT cells, and/or the number of $L$-packets.

\subsection{Cloud embedded in the ISRF}
 
\begin{figure}[h]
\centerline{
\includegraphics[width=4.4cm]{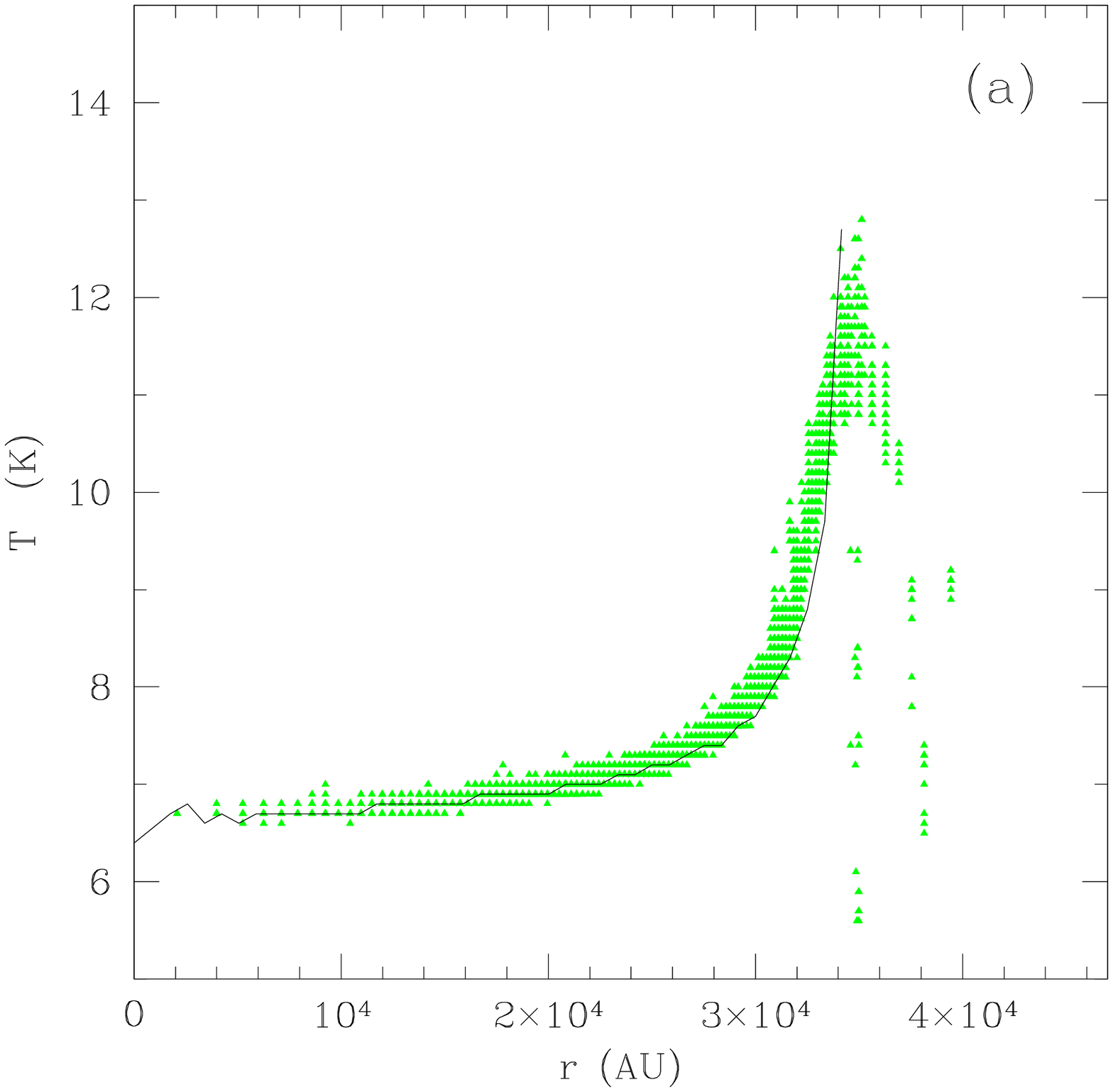}
\includegraphics[width=4.4cm]{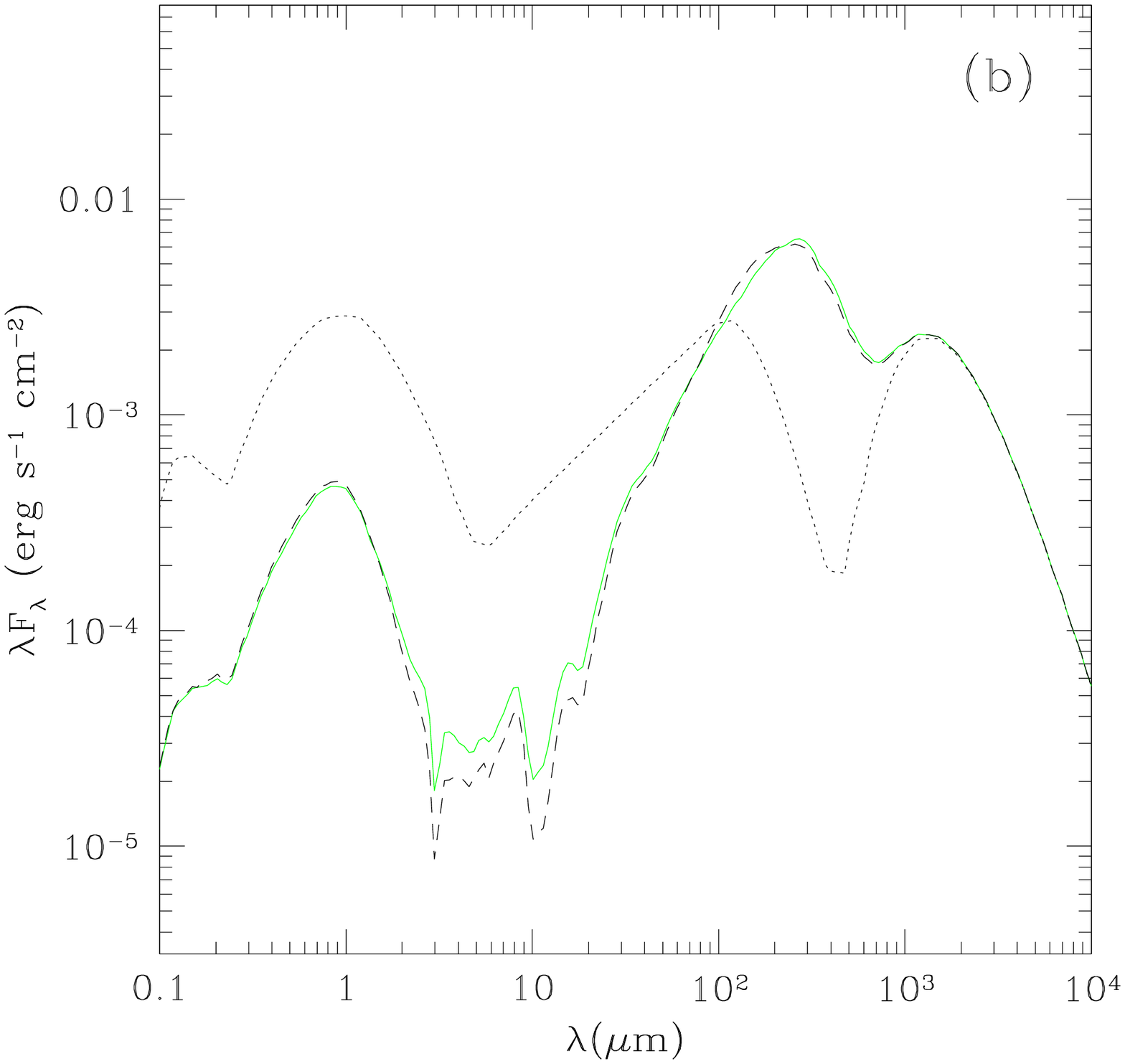}}
\caption{A uniform-density spherical cloud, represented by 200,000 SPH 
particles, is illuminated by the BISRF radiation field, using 
$10^8$ $L$-packets. RT cells are constructed with 
$N_{_{\rm MAX}}=100$ giving $N_{_{\rm CELLS}}=11,862$. {\bf (a)} 
Triangles give the temperatures of the individual RT cells 
and the solid line gives the accurate temperature profile. 
{\bf (b)} The computed SED (solid line) is compared with the 
accurate SED (dashed line) and the illuminating BISRF (dotted 
line).}
\label{fig_sph_bisrf21}
\end{figure}

We use $N_{_{\rm TOTAL}} = 200,000$ SPH particles and set the maximum 
number of particles in an RT cell to $N_{_{\rm MAX}} = 100$, giving 
$N_{_{\rm CELLS}} = 11,862$ RT cells, each containing on average ${\bar N} 
\sim 18$ SPH particles. $10^8$ $L$-packets are then injected from 
random positions on the boundary of the cloud, and with random 
frequencies and injection angles, so as to imitate the isotropic 
Black (1994) interstellar radiation field (BISRF). The advantage of 
this test is that the BISRF covers a wide wavelength range, including 
regions where scattering dominates (UV 
and optical) and regions where absorption/emission dominates (IR 
and submm). Therefore, we can test the validity of the method over 
a wide range of wavelengths. Fig.~\ref{fig_sph_bisrf21} compares 
the resulting temperature profile and SED with accurate results 
obtained using a spherically symmetric grid of thin concentric 
RT cells. The agreement is very good, apart from the irregular RT cells 
at the boundary.

\section{Stars in SPH-RT simulations}
\label{sec:star.grid}

The presence of a normal- or high-temperature star in an SPH 
simulation makes the treatment 
of radiation transport more complicated, firstly because in the 
immediate vicinity of the star the dust is destroyed by sublimation 
and/or chemical sputtering (e.g. Lenzuni et al. 1995), and secondly 
because just  outside this region there are very large temperature 
gradients which cannot be captured by the global grid.

\subsection{Dust destruction radius}
 
The dust destruction temperature is 
estimated to be between $T_{_{\rm DEST}} \sim 800\,{\rm K}$ and 
$T_{_{\rm DEST}} \sim 2100\,{\rm K}$ (Lenzuni et al. 1995; Duschl et 
al. 1996) and depends on the assumed dust composition. The heating 
rate for an isolated dust grain at distance $R$ from a 
star having radius $R_\star$ and surface temperature $T_\star$ is 
\begin{equation}
\label{eq:dust.gain}
\mathcal{G}=\pi r_{\rm d}^2\,
\int_{0}^{\infty}
{Q_\nu \left(\frac{R_\star}{R}\right)^2\,\pi B_\nu (T_{\star})\,d\nu}\, ,
\end{equation}
where $r_{\rm d}$ is the radius of the grain, and $Q_\nu$ is the 
absorption efficiency. Similarly the dust cooling rate is  
\begin{equation}
\label{eq:dust.loss}
\mathcal{L}=
4\pi r_{\rm d}^2\,
\int_{0}^{\infty}Q_\nu\,\pi B_\nu (T)\,d\nu\,,
\end{equation}
where $T$ is the dust temperature. 
Assuming that the dust is in thermal equilibrium, i.e. 
$\mathcal{G}=\mathcal{L}$, and putting $Q_\nu \propto \nu^\beta$ 
($1\stackrel{<}{_\sim}\beta\stackrel{<}{_\sim}2$), we obtain
\begin{equation}
T=\left(\frac{R_\star}{2R}\right)^{2/(4+\beta)}T_\star\,.
\end{equation}
At the dust destruction radius, $T=T_{_{\rm DEST}}$, and hence $R = 
R_{_{\rm DEST}}$, where 
\begin{equation}
R_{_{\rm DEST}}=\frac{R_\star}{2}\, 
\left(\frac{T_\star}{T_{_{\rm DEST}}}\right)^{(4+\beta)/2} \,.
\end{equation}
In the optically thin limit, this radius defines the spherical volume 
around the star which should be devoid of dust. If the envelope of dust 
around the star is optically thick to its own cooling radiation, it will 
act to trap the radiation from the star and $R_{_{\rm DEST}}$ will be 
somewhat larger. The values of $R_{_{\rm DEST}}$ stipulated by Ivezi\'c 
et al. (1997; see below) take this into account.

\begin{figure}
\centerline{
\includegraphics[width=9cm]{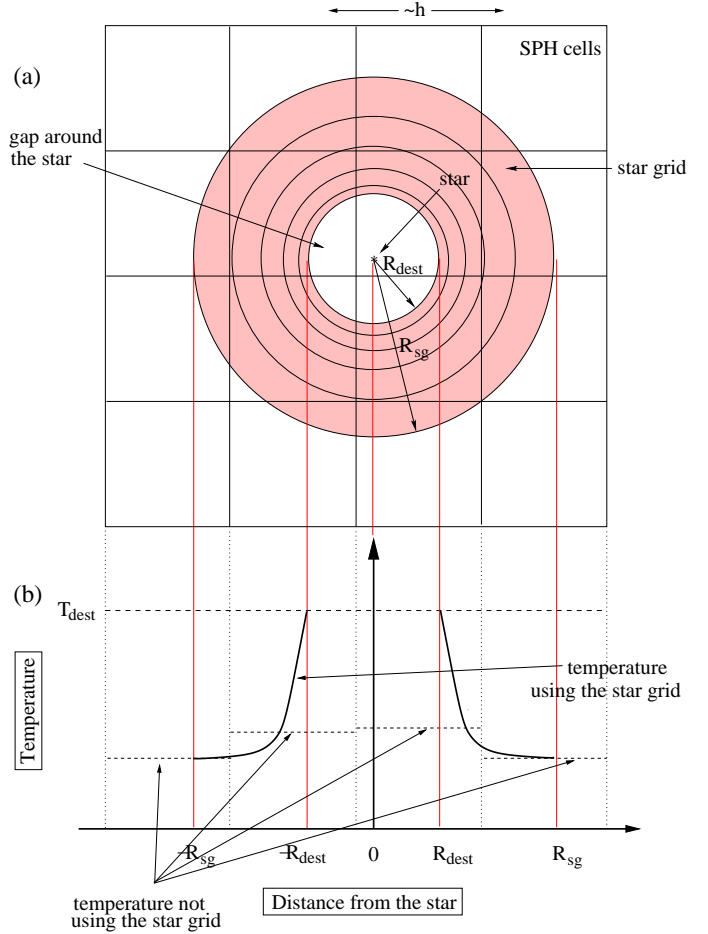}}
\caption{{\bf (a)} The square cells represent the global 
grid derived from the SPH tree, and the concentric circles represent 
the additional star grid. {\bf (b)} The dashed 
horizontal lines show the temperature computed (inaccurately) using 
the global grid on its own, whilst the solid curve shows 
the (accurate) temperature profile computed using the additional 
star grid.}
\label{fig_sph.star.grid}
\end{figure}

\subsection{Steep temperature gradients in the vicinity of a star}
 
The density and temperature gradients close to a star are large, 
and can only be resolved with RT cells whose radial extent is comparable 
with, or smaller than, the dust destruction radius $R_{_{\rm DEST}}$. 
Unless an impractically large number of SPH particles has been used 
in the simulation, the cubic RT cells of the global grid (with size $\sim h$, 
where $h$ is the smoothing length of the SPH kernel) are much larger 
than $R_{_{\rm DEST}}$, as illustrated schematically in 
Fig.~\ref{fig_sph.star.grid}. Thus the steep temperature gradients in 
the vicinity of the star occupy only a few RT cells of the global grid, 
and they are not properly resolved. Consequently the temperatures very 
close to the star are underestimated. (Further out the temperature may be 
overestimated or underestimated, but this is less critical.) The 
upshot is that $L$-packets which are absorbed very close to the star 
are then re-emitted from a wavelength distribution corresponding to 
an inappropriately low temperature, and hence at inappropriately long 
wavelengths. This means that they are able to propagate further away 
from the star, or even to escape from the system altogether. 
Consequently the SED has a deficit of radiation at the short (mainly 
NIR) wavelengths emitted by hot dust, and a corresponding excess of 
radiation at long (mainly FIR) wavelengths. This problem also arises 
in the simulations of Kurosawa et al. (2004).

To resolve this problem, we construct an extra grid of concentric shells 
(a star grid) around each star (see Fig.~\ref{fig_sph.star.grid}). To 
construct a star grid, we must specify:
\begin{itemize}
\item The dust destruction radius, $R_{_{\rm DEST}}$, which is determined 
by the temperature $T_{\star}$ and the radius $R_{\star}$ of the star. 
For simplicity we put
\begin{equation}
\label{eq:lum}
L_{\star} = 
4\pi R_{\star}^2 \sigma T_{\star}^4 = 
L_{\sun}\left(\frac{M_{\star}}{M_{\sun}}\right)^3
+\frac{G M_{\star}\dot{M}_{\star}}{R_{\star}}\,,
\end{equation}
where $M_{\star}$ is the mass of the star and $\dot{M}_{\star}$ is the 
accretion rate onto the star. The first term in Eq.~(\ref{eq:lum}) 
is the intrinsic luminosity of the star, and the second term is the 
accretion luminosity. For a Pre-Main Sequence star we can put 
$R_{\star} \simeq 3 R_{\sun}\,$ (Stahler 1988).

\item The outer radius  $R_{_{\rm SG}}$, of the star grid, 
which should be on the order of the local SPH resolution, or, 
equivalently, on the order of the size of the local cubic RT cells 
of the global grid.

\item The density profile between $R_{_{\rm DEST}}$ and $R_{_{\rm SG}}$, 
for which we adopt
\begin{equation}
\label{eq:sg.dens.profile}
\rho=\rho_0 \left( \frac{R_{_{\rm DEST}}}{R}\right)^p,\,  
R_{_{\rm DEST}}<R<R_{_{\rm SG}}\,,
\end{equation}
with $\rho_0$ fixed by conservation of mass, and $p=0\;{\rm to}\;2\,$.
\end{itemize}

Once all the above parameters have been specified, the star grid 
around each star is constructed by dividing up the space 
between $R_{_{\rm DEST}}$ and $R_{_{\rm SG}}$ into concentric RT cells of 
equal logarithmic width; typically $\sim 30$ RT cells are sufficient 
for each star grid. The complication that this 
introduces into an RT calculation is minimal. There is some overlap 
between the two sets of RT cells (the cubic cells of the global grid 
and the spherical cells of the star grid), 
but this does not create any major problems, as our 
tests in the next section demonstrate. $L$-packets which are 
inside the star grid interact with the RT cells 
of that grid, rather than the RT cells of the global grid. 

However, we note that a spherically symmetric star grid may not be appropriate 
when the dust close to a star is concentrated in a disc. This situation 
can be handled with a simple extension to the method we have presented here, 
and it will be discussed in the companion paper (Stamatellos et al. 2005).

\section{Ivezi\'c test}
We test the above method for including stars in radiative transfer 
calculations within SPH, using the configuration proposed by 
Ivezi\'c et al. (1997), {\it viz.} a star with $T_\star=2500\,{\rm K}$ 
surrounded by a spherical 
envelope with an $R^{-2}$ density profile. We perform the test 
for envelopes having visual radial optical depths (from the centre 
to the edge of the envelope) $\tau_{\rm V}=1$ and $\tau_{\rm V}=100$
(see online appendix). 
For $\tau_{\rm V}=1$, we set $R_{_{\rm DEST}} = 9.11 R_\star$; and for 
$\tau_{\rm V}=100$, $R_{_{\rm DEST}} = 17.67 R_\star$, as stipulated by 
Ivezi\'c et al. (1997). For both cases we set $R_{_{\rm SG}} = 20 
R_{_{\rm DEST}}$ and $R_{_{\rm OUTER}} = 1000 R_{_{\rm DEST}}$ (where 
$R_{_{\rm OUTER}}$ is the outer radius of the envelope). Note that we 
do not need to specify $R_\star$.

We represent the envelope using 20,000 SPH particles, distributed 
randomly so that they approximate to an 
$R^{-2}$ profile. (The approximation could be improved by settling 
the distribution to get rid of Poisson fluctuations, but we have not 
done this.) We  use the SPH tree to construct a global grid with 
$N_{_{\rm MAX}}=55$ (which gives $N_{_{\rm CELLS}} = 1276$ and ${\bar N} 
= 15$). In addition, we have the option to introduce a star grid 
comprising 30 concentric RT cells, between 
$R_{_{\rm DEST}}$ and $R_{_{\rm SG}}$.

The results are compared with the very accurate results obtained 
using a one-dimensional spherically symmetric code in Stamatellos 
\& Whitworth (2003), hereafter `the benchmark calculation'.

\begin{figure}
\centerline{
\includegraphics[width=4.4cm]{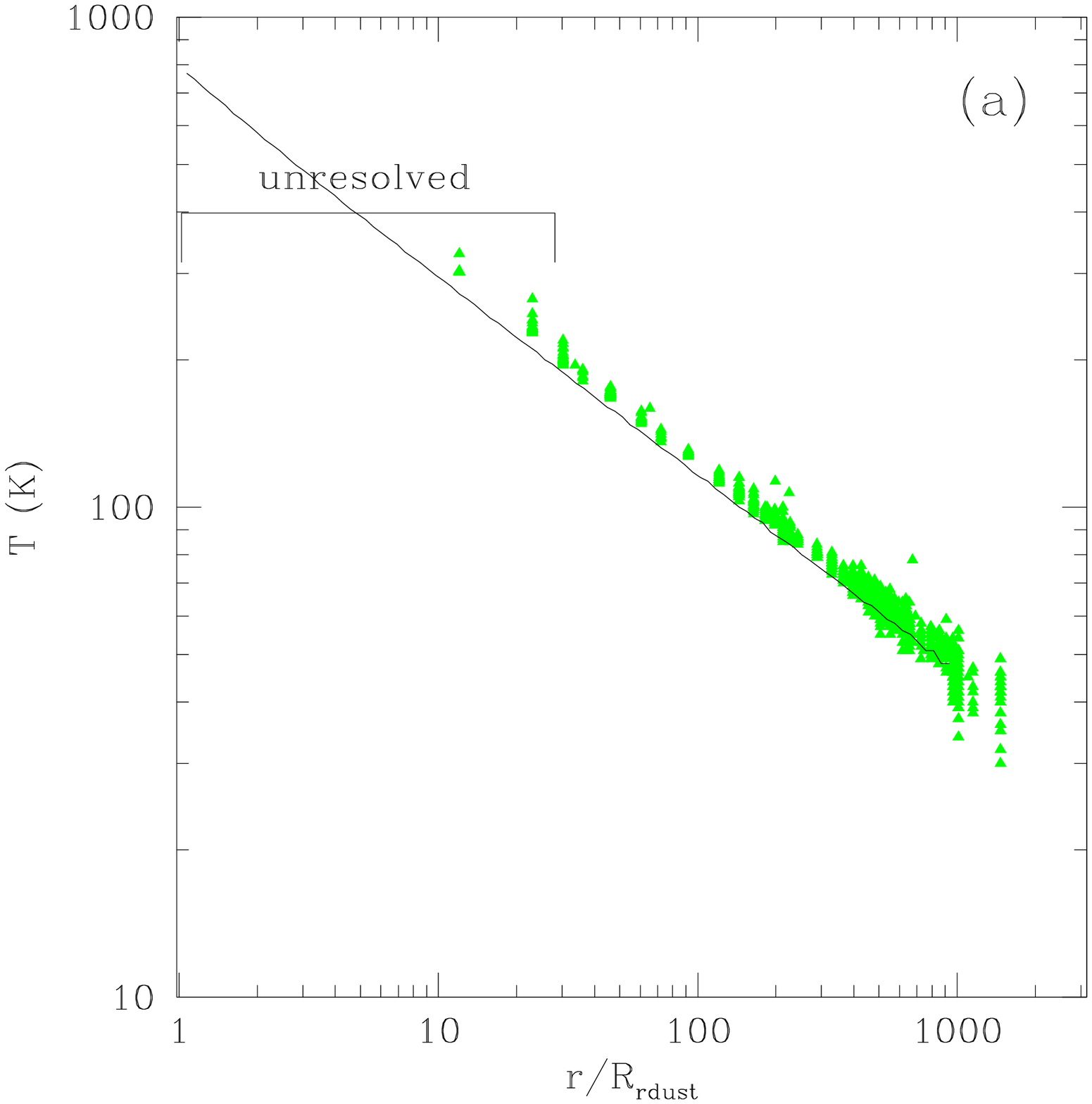}
\includegraphics[width=4.4cm]{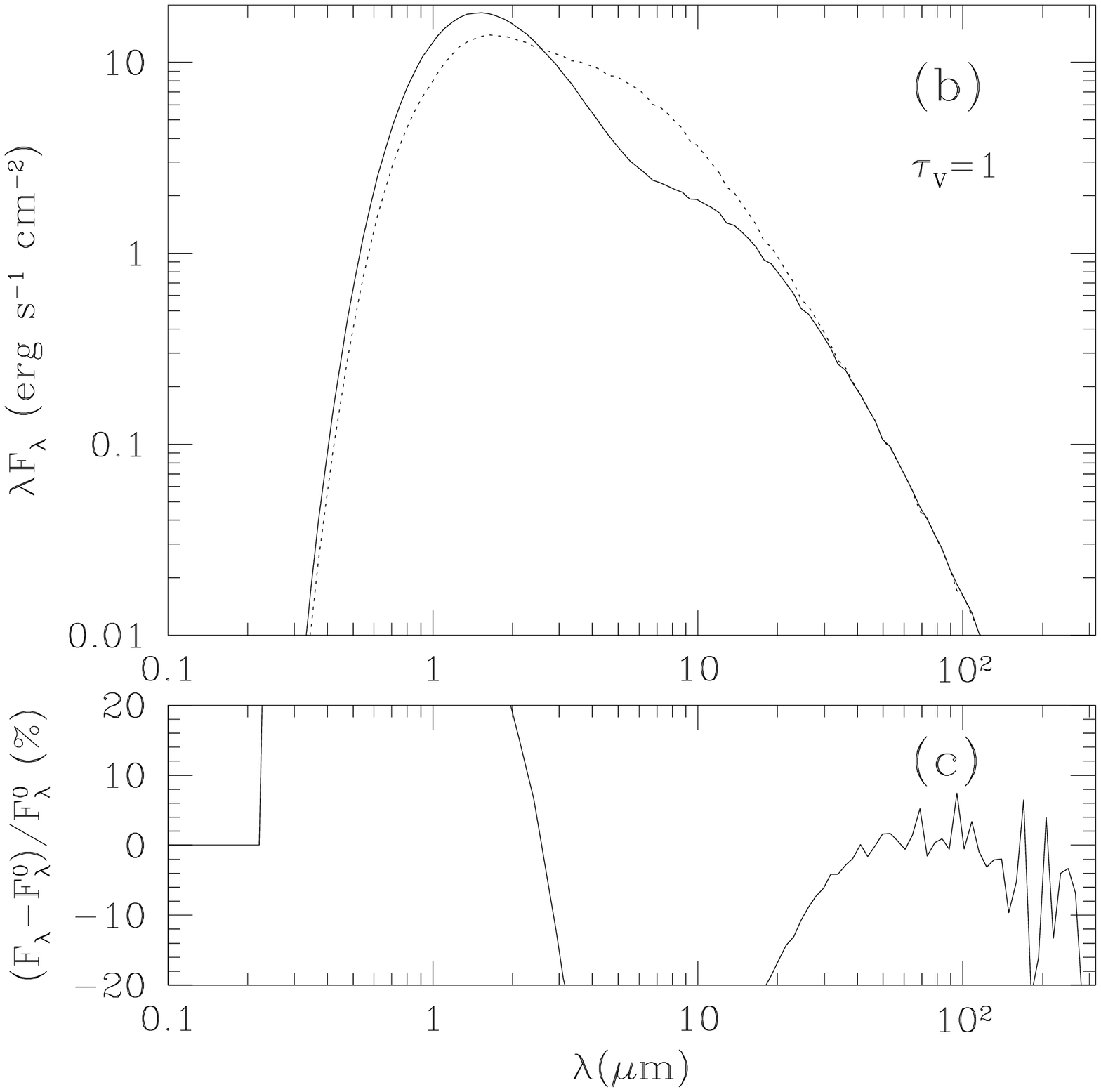}}
\caption{Ivezi\'c test with $\tau_{_{\rm V}}=1$, using $5\times10^7$ 
$L$-packets. The envelope is represented by 20,000 SPH particles and a 
global grid of $N_{_{\rm CELLS}} = 1,276$ RT cells (with ${\bar N} = 
15$), but no additional star grid around the star. {\bf (a)} 
The triangles represent the temperatures of the individual RT cells 
plotted against radius, whilst the solid line represents the 
benchmark calculation. The global grid on its own fails to 
capture the temperature profile close to the star. {\bf (b)} The solid 
line shows the SED computed here, and the dotted line shows the 
benchmark calculation. {\bf (c)} The percentage difference between the 
flux computed here and the benchmark calculation.}  
\label{fig_sph_iv1_wrong}
\centerline{
\includegraphics[width=4.4cm]{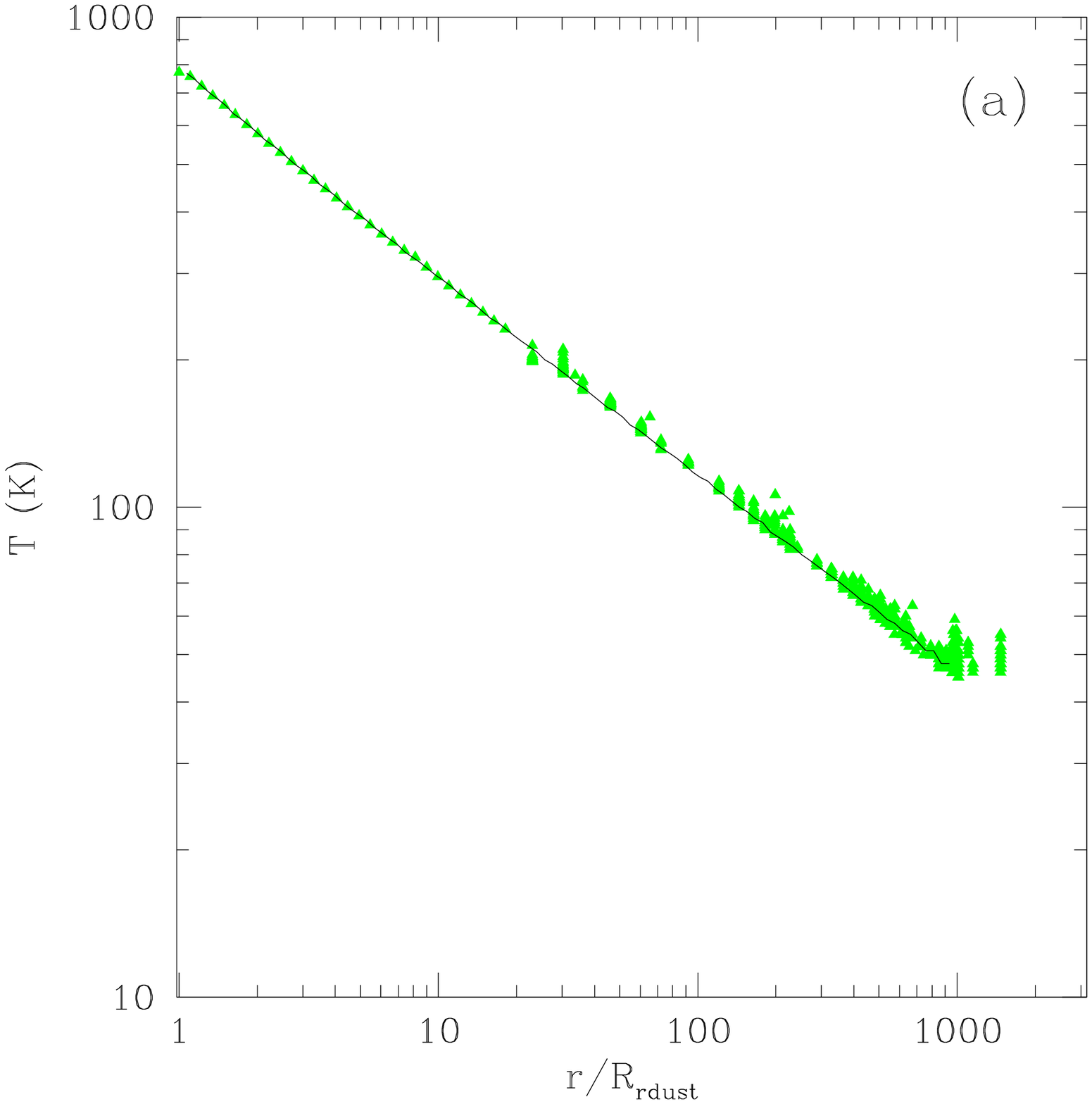}
\includegraphics[width=4.4cm]{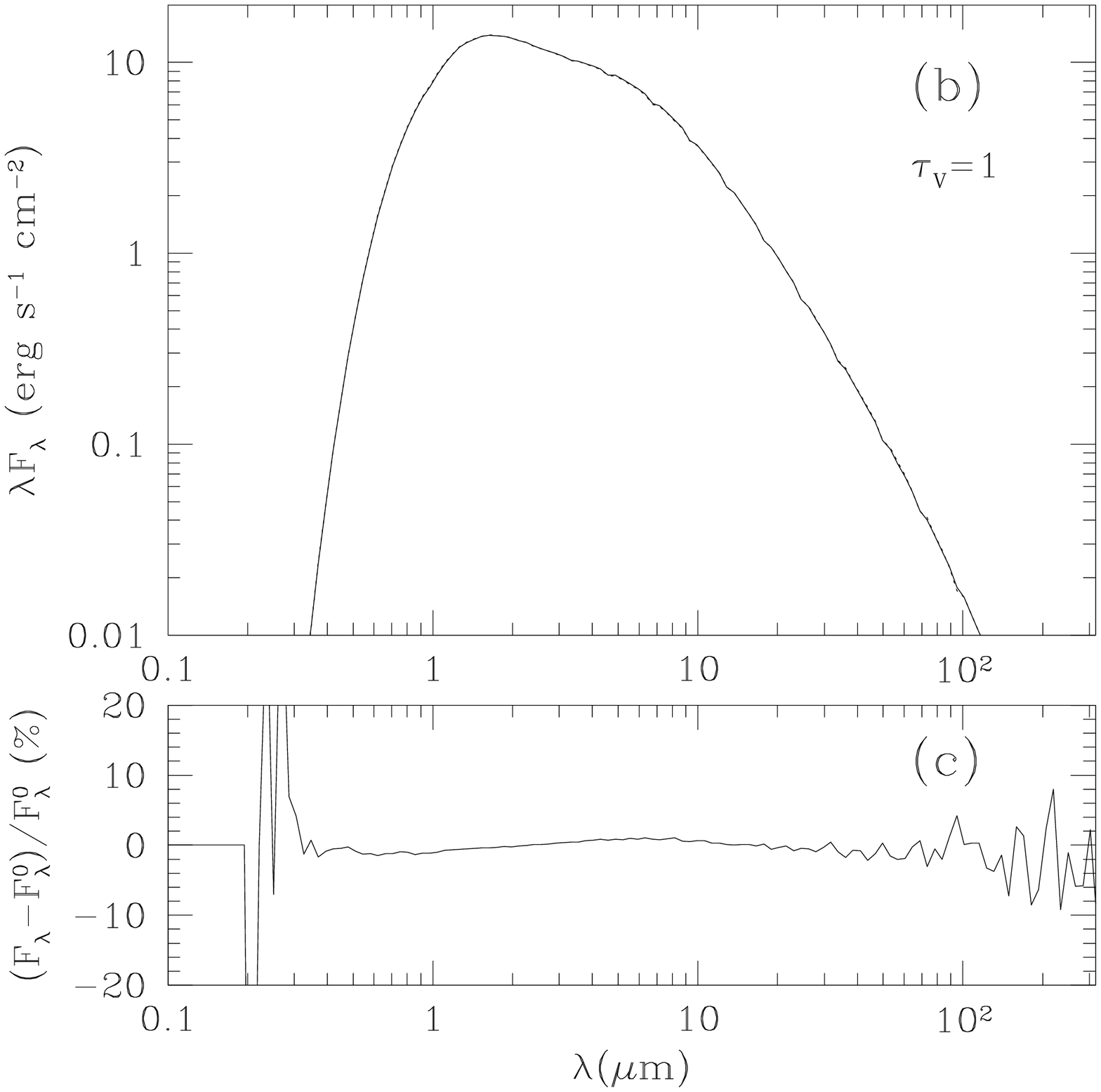}}
\caption{Same as Fig.~\ref{fig_sph_iv1_wrong} but with an additional 
star grid around the star, i.e. a volume devoid
of matter inside $R_{_{\rm DEST}}$ and 30 concentric spherical RT cells 
between $R_{_{\rm DEST}}$ and $R_{_{\rm SG}} = 20 R_{_{\rm DEST}}$. Agreement 
with the benchmark calculation is now very good. Small differences 
are due to statistical noise.}
\label{fig_sph_iv1}
\end{figure}

We have performed the Ivezi\'c test with $\tau_{\rm V} = 1$ 
and $p = -2 $, using $5\times 10^7$ $L$-packets, both without 
(Fig.~\ref{fig_sph_iv1_wrong}), and with (Fig.~\ref{fig_sph_iv1}), a 
star grid around the central star. 

Without a star grid around the central star 
(Fig.~\ref{fig_sph_iv1_wrong}) the cubic RT cells of the global grid 
in the vicinity of the star have size $\sim 20R_{_{\rm DEST}}$, and 
therefore they are unable to capture the large temperatures and densities 
there. This has two effects. First, the hottest dust is only 
at $\sim 300\,{\rm K}$ (rather than $\sim 800\,{\rm K}$ in the benchmark 
calculation), and consequently the SED shows a deficit of radiation at 
wavelengths around $8\,\mu{\rm m}$. Second, the uniform density in the 
individual RT cells reduces the net radial  optical depth, 
allowing more radiation to escape directly from the central star 
(specifically $\tau_{\rm V}^{^{\rm EFFECTIVE}} \sim 0.3$); this shows up in 
the SED as an excess of radiation at wavelengths around $1\,\mu{\rm m}$. 

With a star grid around the central star 
(Fig.~\ref{fig_sph_iv1}), the innermost RT cell (immediately outside 
$R_{_{\rm DEST}}$) has radial extent $\sim 0.1 
R_{_{\rm DEST}}$, and the high temperatures and densities near the star are 
properly represented. Agreement with the benchmark calculation is then 
very good. Small differences are due to statistical noise in RT cells which 
do not intercept many $L$-packets.

\section{Discussion}
 
In this paper we have presented a method for performing radiative 
transfer calculations on individual time-frames from SPH 
simulations. We use the SPH tree to construct a global grid 
of radiative transfer (RT) cells. This grid is constructed so that 
each RT cell contains a number of SPH particles close to the number 
of neighbours that each particle must have in SPH for smoothing 
purposes. The method thereby creates RT cells with size on the 
order of the local SPH smoothing length.

We test the method using the thermodynamic equilibrium test for a 
uniform-density spherical cloud, bathed in a blackbody 
radiation field. The method performs this test well. The global grid 
of RT cells that represent the cloud acquire the same temperature as the 
blackbody radiation field. The temperature is slightly underestimated near the 
boundary of the cloud due to the presence of RT cells that contain only a few 
SPH particles (irregular RT cells). The method also performs well for a 
uniform-density cloud illuminated by the interstellar radiation field.

We then examine a system comprising a low-temperature star embedded at 
the centre of a uniform-density spherical cloud. The temperature field 
and the SED are calculated with good accuracy. Very close to the star  
the temperature is slightly underestimated due to the fact that the 
temperature field there is not properly resolved. This in turn distorts 
the SED of the system slightly at short wavelengths. The problem is more 
pronounced when a normal- or high-temperature star is embedded in the 
cloud. The region very close to the star is then characterised by a very 
steep temperature gradient which cannot normally be captured by the 
cubic RT cells of the global grid, because these RT cells are too large.

To solve this problem we introduce an additional grid of concentric 
spherical RT cells around the star (the star grid). The 
star grid accounts for the dust-free region near the star, 
and its concentric spherical RT cells are small enough to capture steep 
temperature gradients, so that temperatures are calculated correctly 
close to the star. We test the use of a global grid in tandem 
with a star grid around the star, using the configuration 
proposed by Ivezi\'c et al. (1997), and show  that it performs very 
well, even if the optical depth is very large.

Our radiative transfer code can therefore be applied to arbitrary SPH 
density fields to produce dust temperature distributions, SEDs and 
isophotal maps at different wavelengths. These can then be used to 
compare the results of hydrodynamic simulations with observations. In 
the companion paper (Stamatellos et al.~2005) we apply this method to 
the collapse of turbulent molecular cores and the early stages of star 
formation.  

\begin{acknowledgements}

We  gratefully acknowledge support from the EC Research Training 
Network ``The Formation and Evolution of Young Stellar Clusters'' 
(HPRN-CT-2000-00155) and from PPARC (PPA/G/O/2002/00497). We would 
also like to thank  J. Barnes for using his tree code (available 
online at {\texttt http://www.ifa.hawaii.edu/$\sim$barnes/software.html}).
\end{acknowledgements}

\Online
\appendix 
\section{Supplementary tests}
\subsection{Low-temperature star surrounded by a spherical envelope}
 
\begin{figure}[hb] 
\centerline{
\includegraphics[width=4.4cm]{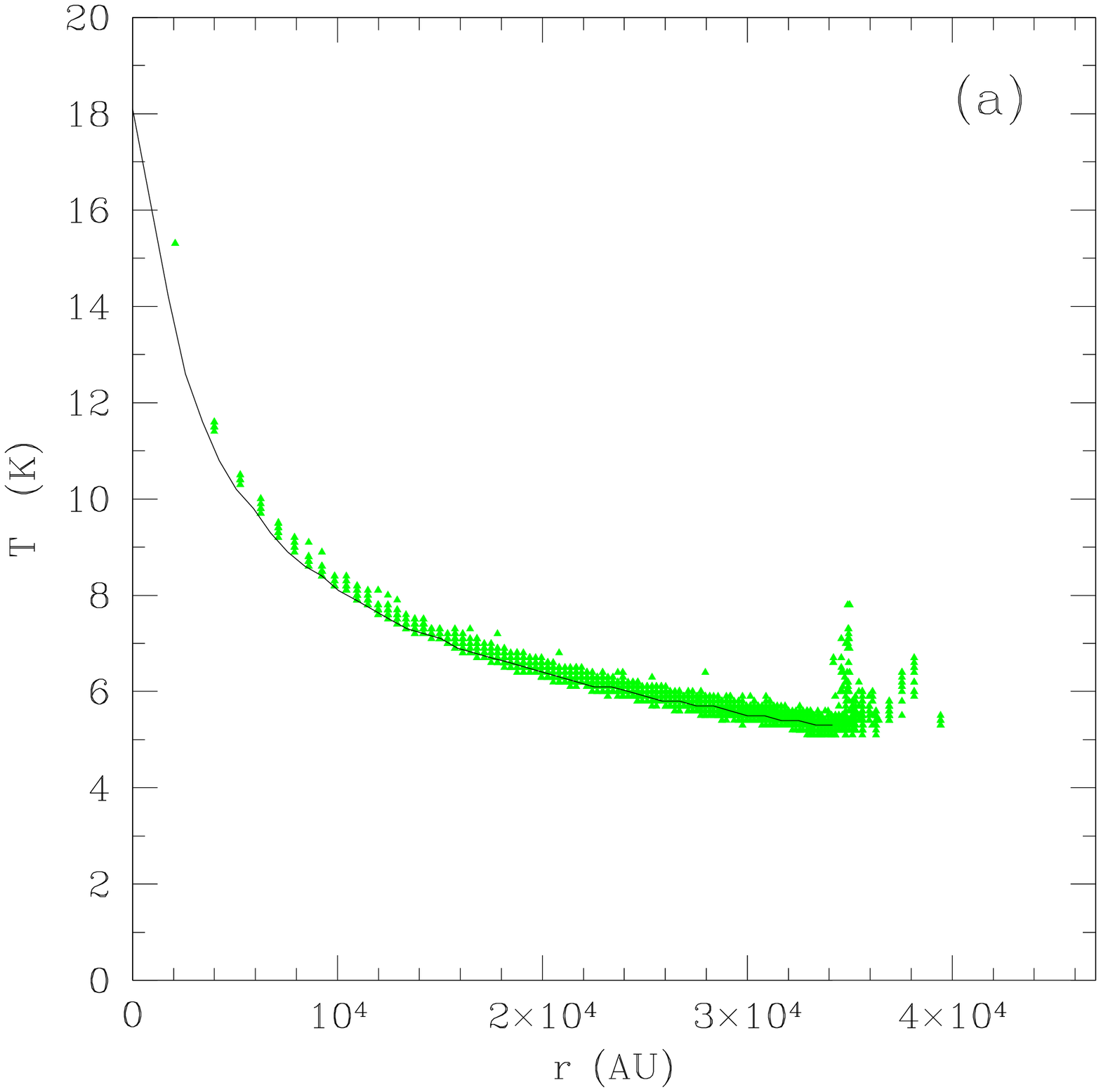}
\includegraphics[width=4.4cm]{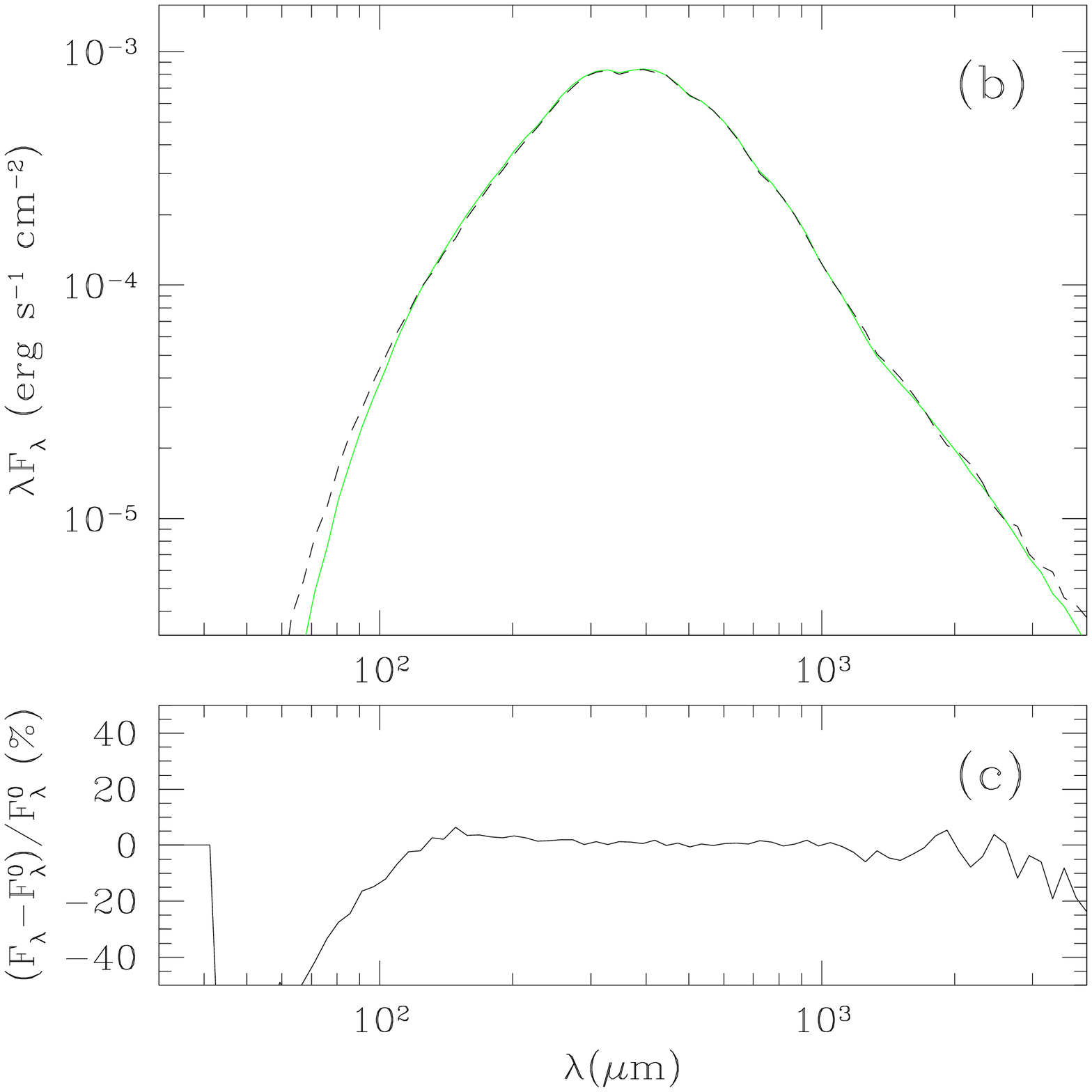}}
\caption{A uniform-density spherical cloud, represented by 200,000 SPH 
particles, is heated by a central low-temperature star with $L_* = 
7 {\rm L}_\odot$ and $T_* = 30\,{\rm K}$, which emits $10^8$ 
$L$-packets. RT cells are constructed with $N_{_{\rm MAX}}=60$, giving 
$N_{_{\rm CELLS}}=13,050$. {\bf (a)} Triangles give the temperatures 
of the individual RT cells and the solid line gives the accurate 
temperature profile. {\bf (b)} The computed SED (solid line) is 
compared with the accurate SED (dashed line). {\bf (c)} The 
precentage error in the SED.}
\label{fig_sph_cs.29}
\end{figure}

We use $N_{_{\rm TOTAL}} = 200,000$ SPH particles and set the maximum 
number of particles in an RT cell to $N_{_{\rm MAX}} = 60$, giving 
$N_{_{\rm CELLS}} = 13,050$ cells, each containing on average ${\bar N} 
\sim 16$ SPH particles. $10^8$ $L$-packets are emitted by a central 
low-temperature `star' having luminosity $L_* = 7 {\rm L}_\odot$ and 
surface temperature 
$T_\star = 30\,{\rm K}$. Fig.~\ref{fig_sph_cs.29} compares the resulting 
temperature profile and SED with accurate results obtained using a 
spherically symmetric grid of thin concentric RT cells. The global 
grid used here has difficulty capturing the steep temperature 
gradient near the star. As a result the temperatures near the 
star are underestimated and the SED shows a deficit at 
short wavelengths. The temperature is also not very accurate near the 
boundary, due to the irregular RT cells there; and there are some small 
uncertainties in the SED at wavelengths longer than $2000~\micron$, 
due to the small number of $L$-packets generated at these wavelengths. 
Otherwise the agreement is good.

\subsection{Ivezi\'c test with $\tau_{\rm V}=100$ and $p = - 2$}
 
\begin{figure}[h]
\centerline{
\includegraphics[width=4.4cm]{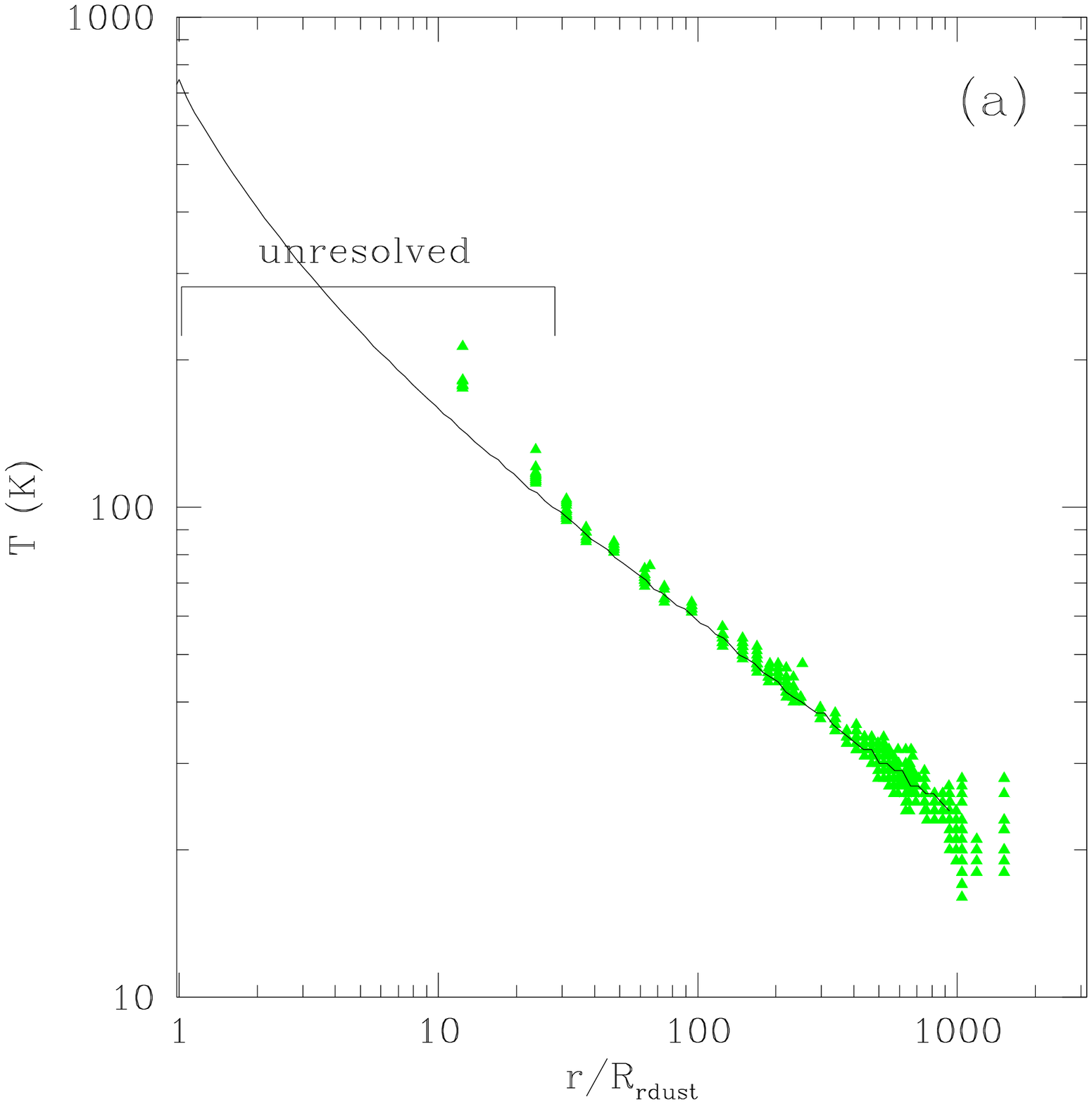}
\includegraphics[width=4.4cm]{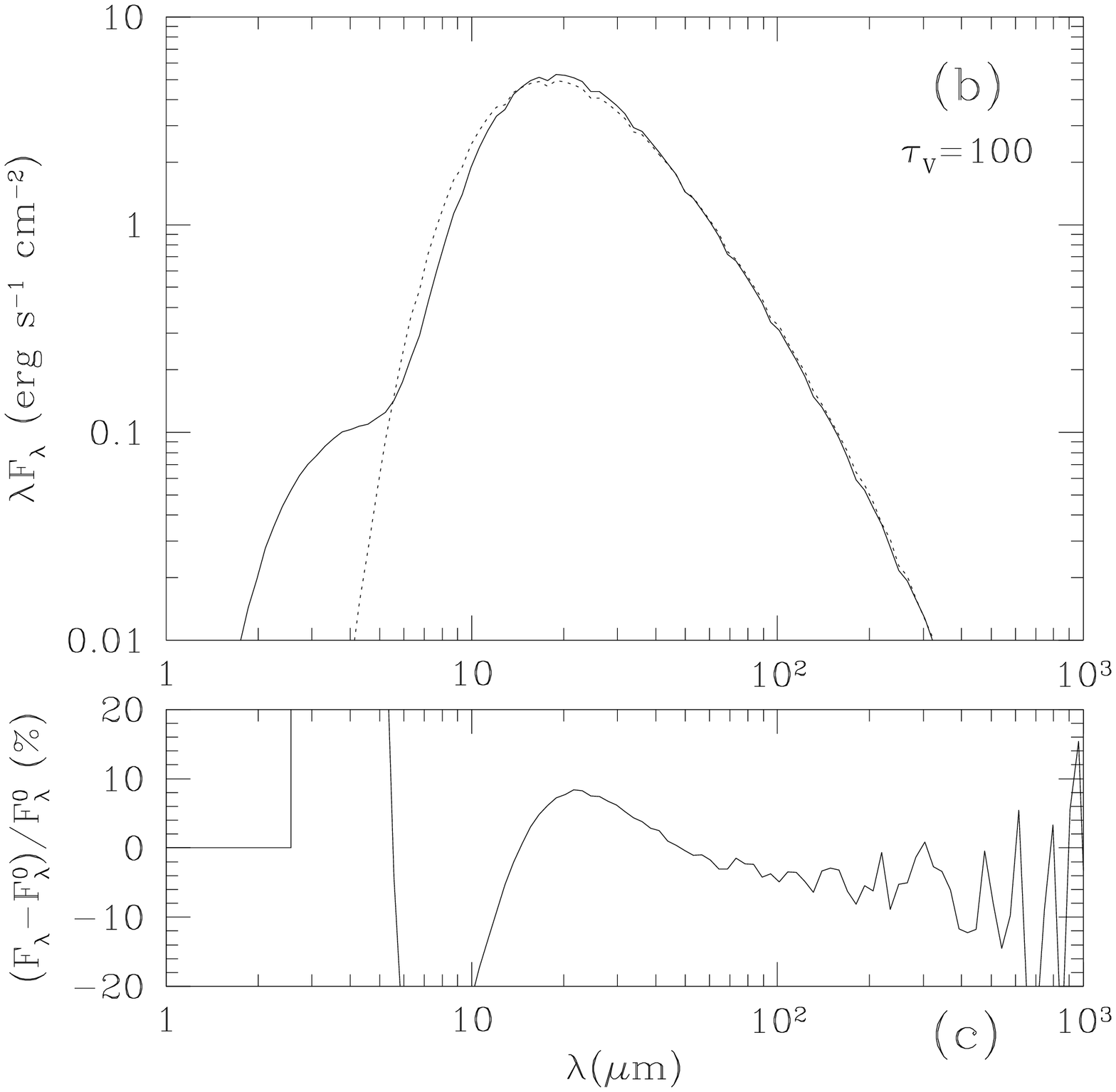}}
\caption{Ivezi\'c test with $\tau_{\rm V}=100$, using $10^7$ 
$L$-packets. The envelope is represented by 20,000 SPH particles and a 
global grid of $N_{_{\rm CELLS}} = 1,310$ RT cells (with ${\bar N} = 
15$), but no additional star grid around the star. {\bf (a)} 
The triangles represent the temperatures of the individual 
RT cells plotted against radius, whilst the solid line represents the 
benchmark calculation. The global grid on its own fails to 
capture the temperature profile close to the star. {\bf (b)} The solid 
line shows the SED computed here, and the dotted line shows the 
benchmark calculation. {\bf (c)} The percentage difference between the 
flux computed here and the benchmark calculation.}
\label{fig_sph_iv100_wrong}
\end{figure}

\begin{figure}[h]
\centerline{
\includegraphics[width=4.4cm]{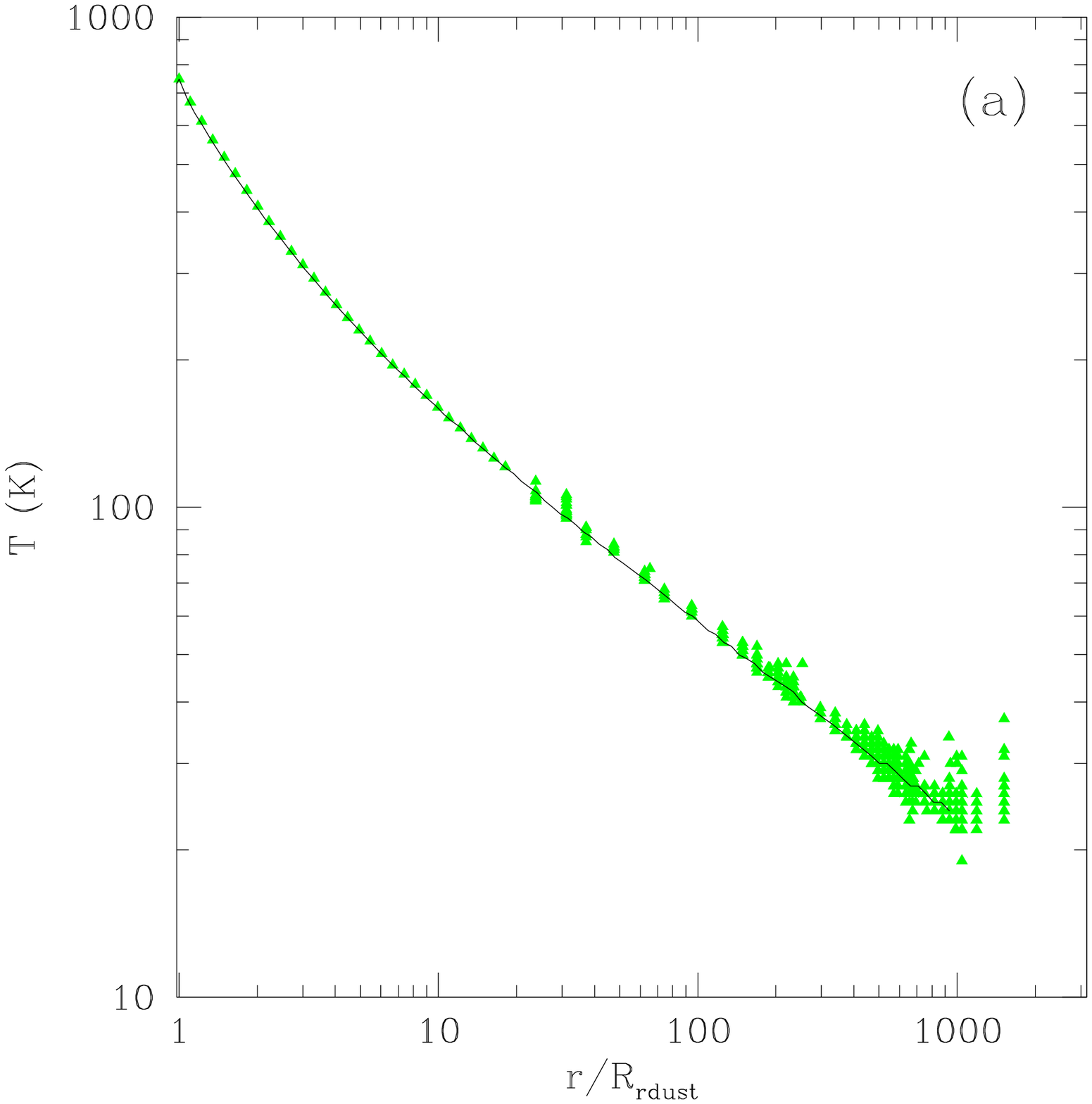}
\includegraphics[width=4.4cm]{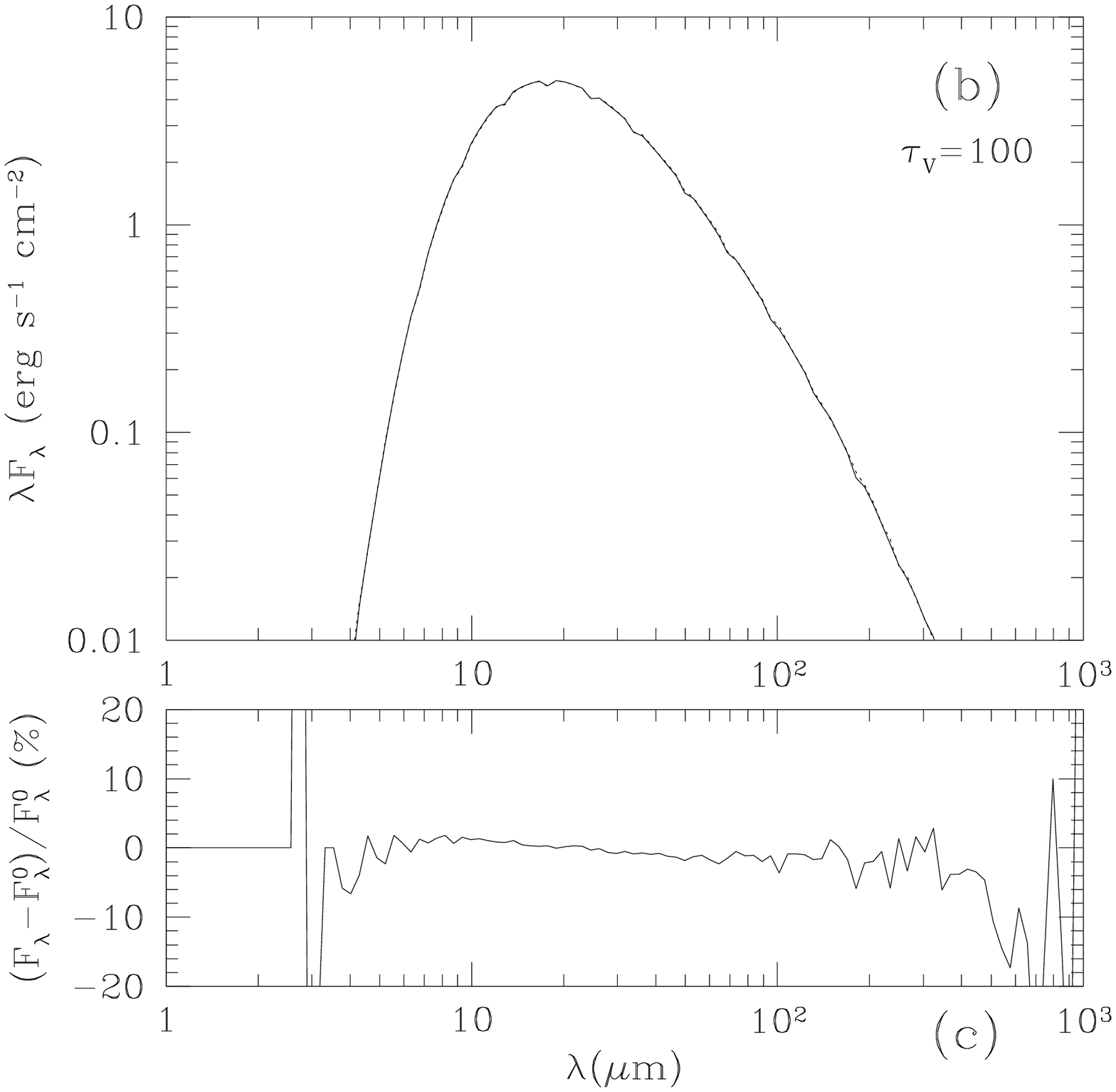}}
\caption{Same as Fig.~\ref{fig_sph_iv100_wrong} but with an additional 
star grid around the star, i.e. a volume devoid
of matter inside $R_{_{\rm DEST}}$ and 30 concentric spherical RT cells 
between $R_{_{\rm DEST}}$ and $R_{_{\rm SG}} = 20 R_{_{\rm DEST}}$. Agreement 
with the benchmark calculation is now very good. Small differences 
are due to statistical noise.}
\label{fig_sph_iv100}
\end{figure}

We have  performed the Ivezi\'c test with $\tau_{\rm V} = 100$ 
and $p = -\,2 $, using $10^7$ $L$-packets, both without 
(Fig.~\ref{fig_sph_iv100_wrong}), and with (Fig.~\ref{fig_sph_iv100}), 
a star grid around the central star.

Without a star grid around the central star (Fig.~\ref{fig_sph_iv100_wrong}), 
the cubic RT cells of the global grid in the 
vicinity of the star are again unable to capture the large temperatures 
and densities there. Consequently the hottest dust is only at 
$\sim 200\,{\rm K}$ (rather than $\sim 800\,{\rm K}$ in the benchmark 
calculation). The SED is dominated by the emission from this 
dust, which peaks around $20\,\mu{\rm m}$. In addition, the uniform 
densities in the RT cells reduces the  effective 
radial optical depth ($\tau_{\rm V}^{^{\rm EFFECTIVE}} \sim 20$) and 
so there is a small shoulder at $\sim 3\,\mu{\rm m}$, due to radiation 
which escapes directly from the central star.

With a star grid around the central star 
(Fig.~\ref{fig_sph_iv100}) agreement with the benchmark calculation is 
very good. Small differences are again  due to statistical noise. We 
emphasize that these calculations were performed with only $10^7$ 
$L$-packets.


\end{document}